# Molecular Beam Epitaxy Growth of Tetragonal FeS Film on SrTiO$_3$(001) Substrate

Kun Zhao(赵琨)[1], Haicheng Lin(林海城)[1], Wantong Huang(黄万通)[1], Xiaopeng Hu(胡小鹏)[1,2], Xi Chen(陈曦)[1,2], Qi-Kun Xue(薛其坤)[1,2], Shuai-Hua Ji(季帅华)[1,2*]

[1] State Key Laboratory of Low-Dimensional Quantum Physics, Department of Physics, Tsinghua University, Beijing 100084, China

[2] Collaborative Innovation Center of Quantum Matter, Beijing, China

* To whom correspondence should be addressed. Email: shji@mail.tsinghua.edu.cn

**Abstract** We report the successful growth of tetragonal FeS film with one or two unit-cell (UC) thickness on SrTiO$_3$(001) substrate by molecular beam epitaxy. Large lattice constant mismatch with the substrate leads to high density of defects in single UC FeS, while it has been significantly reduced in double UC thick film due to the lattice relaxation. The scanning tunneling spectra on the surface of FeS thin film reveal the electronic doping effect of single UC FeS from the substrate. In addition, at the Fermi level, the energy gaps of approximate 1.5 meV are observed in films of both thicknesses at 4.6 K and below. The absence of coherence peaks of gap spectra may be related to the preformed Cooper-pairs without phase coherence.

The discovery of interface-enhanced high-temperature superconductivity in FeSe/SrTiO$_3$(001) has attracted tremendous attention recently. Single unit-cell (UC) FeSe film grown on SrTiO$_3$(001) substrate has a superconducting gap of about 20 meV and the superconducting transition temperature may be as high as 109 K [1, 2], which is extremely enhanced compared with that of bulk FeSe [3, 4]. However, both single UC FeSe film grown on graphene substrate and double UC FeSe film grown on SrTiO$_3$(001) substrate are non-superconducting [1, 5, 6]. This intriguing observation has immediately inspired an impressive amount of experimental and theoretical studies in order to clarify the hidden physical mechanism of this novel high-temperature superconductivity [7-20]. Now, it has already been widely believed that it originates from interfacial superconductivity, in which the interface charge transfer and interface-enhanced electron-phonon coupling play essential roles together [21].

The interface-enhanced high-temperature superconductivity has also been observed in many other interfacial material systems [22-25]. However, so far, most of these interfacial material systems are combined single UC FeSe film with various substrates. If we combine SrTiO$_3$(001) substrate with other single UC film instead of FeSe, it is natural to ask whether the interface-enhanced high-temperature superconductivity exists in this interfacial material system or not. This study can help us to further understand the origin of interface-enhanced high-temperature superconductivity but also broaden the way to search for more superconducting interfacial material systems.

Bulk tetragonal FeS has very similar lattice structure and band structure with FeSe and is also superconducting with a superconducting gap of about 1 meV and the superconducting transition temperature is 4.5 K [26-32]. In this paper, we successfully grow tetragonal FeS film on SrTiO$_3$(001) substrate by molecular beam epitaxy (MBE). Combining low-temperature growth and high-temperature post-annealing in vacuum, sample quality has been significantly improved. The lattice structures and electronic structures of single UC and double UC FeS are revealed by in situ ultra-low-temperature scanning tunneling microscopy (STM). Single UC FeS has a strong interaction between SrTiO$_3$(001) substrate and an obvious electronic doping from the substrate. An energy gap of approximate 1.5 meV in the vicinity of the Fermi level are observed in FeS films of both thicknesses.

The experiments were performed on a MBE-STM combined system with a base pressure of 1 × 10$^{-10}$ Torr (Unisoku). The 0.05wt% Nb-doped SrTiO$_3$(001) substrates were degassed at 600 °C for 3 hours, and then annealed at 1200 °C for 20 minutes to get clean and flat TiO$_2$-terminated surfaces. Atomic S flux was supplied by FeS thermal decomposition due to S has a very low vapor pressure. FeS films were grown by co-evaporating high-purity Fe and FeS from standard Knudsen cells. The growth was set under a S-rich condition and monitored by an *in situ* reflection high-energy electron diffraction (RHEED). The growth rate was about 0.17 UC per minute. After post-annealing in vacuum to further improve the quality of FeS film, the sample was immediately transferred into the *in situ* STM head to perform local topographic imaging and scanning tunneling spectroscopy (STS) measurements. All the STM topographic images were acquired at 4.6 K with a polycrystalline Pt-Ir alloy tip, which was modified and calibrated on the clean Ag(111) surface. The d$I$/d$V$ spectra were acquired at 4.6 K or 60 mK by the standard lock-in technique.

Tetragonal FeS has a similar lattice structure ($a$ = 3.68 Å, $c$ = 5.03 Å) [27] with PbO-type FeSe ($a$ = 3.77 Å, $c$ = 5.48 Å) [3] but with smaller lattice constants. Tetragonal FeS is a metastable phase comparing with more stable hexagonal phases, which makes it quite challenging for the MBE growth. In the growth progress, it always shows mixture of both phases under most conditions, which limits the quality of epitaxial tetragonal FeS film. We introduce SrTiO$_3$(001) substrates with square lattice on the surface to obtain the pure tetragonal phase.

We investigated the effects of substrate temperature on the MBE growth. Fig. 1a, c, e show the STM topographic images of epitaxial FeS films grown at substrate temperatures of 250 °C, 400 °C and 450 °C, respectively, and Fig. 1b, d, f show the corresponding atomic-resolved images. At a low substrate temperature of 250 °C, the epitaxial FeS film has a better morphology, but the quality of crystallization is relatively lower. When the substrate temperature is raised to 400 °C, the epitaxial FeS film begins to decompose and the morphology becomes poorer while the quality of crystallization becomes better and the tetragonal lattice of FeS film can be clearly observed in the atomic-resolved images (Fig. 1d). With further increasing of the substrate temperature to 450 °C, the surface morphology becomes better again, while the epitaxial FeS film is totally changed to the more stable hexagonal phase. Therefore, the quality of as-grown FeS film is not high enough and needs to be improved by post-annealing in vacuum. Only using the combined method of low-temperature growth and high-temperature post-annealing in vacuum can the high-quality epitaxial tetragonal FeS film be obtained. The optimal substrate temperature for the tetragonal FeS film growth is 250 °C.

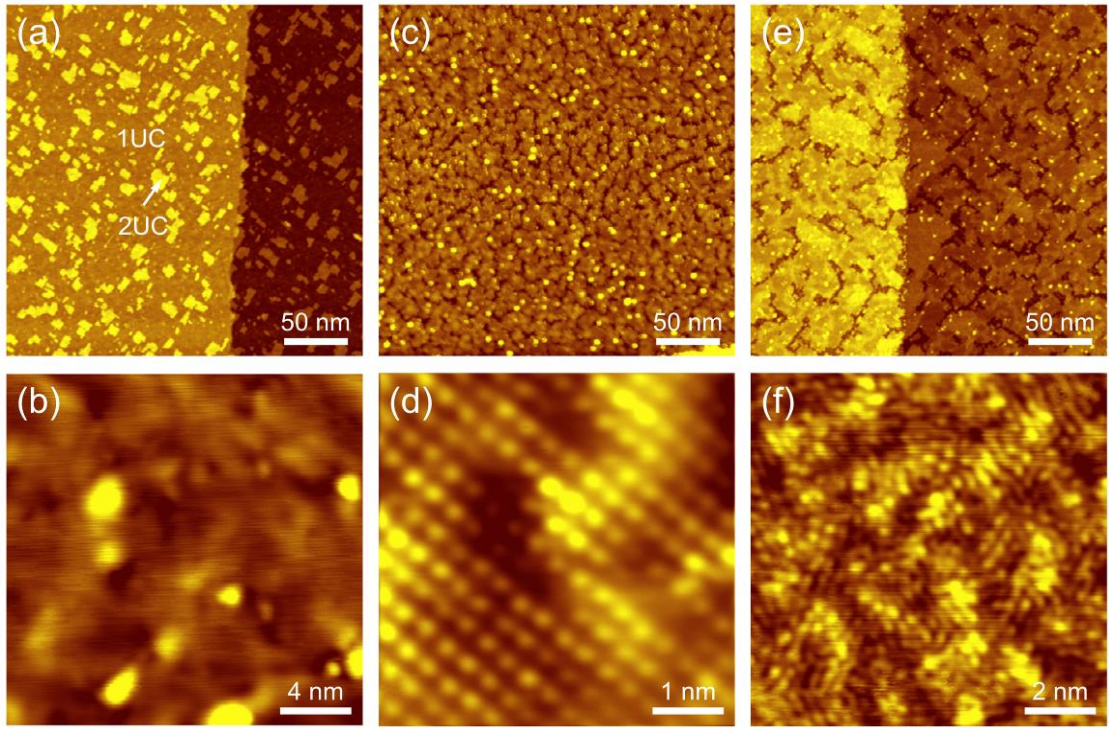

Fig. 1. The effects of substrate temperature on the MBE growth of tetragonal FeS film. (a), (c), (e) The STM topographic images of epitaxial FeS films grown at substrate temperatures of (a) 250 °C, (c) 400 °C and (e) 450 °C, respectively (set point: sample bias $V_s$ = 3.0 V, tunneling current $I_t$ = 20

pA). (b), (d), (f) The corresponding atomic-resolved images of (a), (c), (e) (set point: (b) $V_s$ = 3.0 V, $I_t$ = 20 pA, (d) $V_s$ = 1.0 V, $I_t$ = 100 pA, (f) $V_s$ = 0.5 V, $I_t$ = 100 pA).

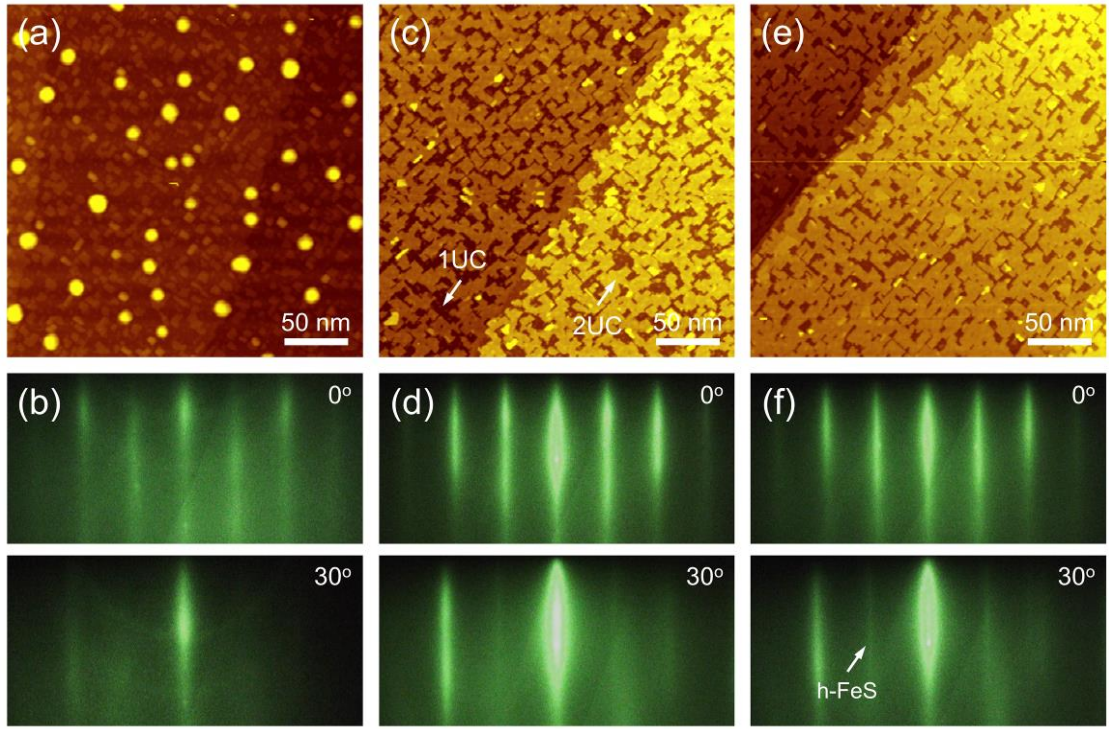

Fig. 2. The effects of FeS source temperature on the MBE growth of tetragonal FeS film. (a), (c), (e) The STM topographic images of epitaxial FeS films grown at FeS source temperatures of (a) 770 °C, (c) 780 °C and (e) 795 °C, respectively (set point: $V_s$ = 3.0 V, $I_t$ = 20 pA). (b), (d), (f) The corresponding RHEED patterns of (a), (c), (e).

Another parameter that affects the MBE growth is the flux ratio of Fe and S. We then investigated the effects of different flux rates of S on the epitaxial FeS film growth with the same flux rate of Fe source. Figure 2a, c, e show the STM topographic images of epitaxial FeS films grown with FeS source temperatures of 770 °C, 780 °C and 795 °C, respectively, and Figure 2b, d, f show the corresponding RHEED patterns. For a low FeS source temperature of 770 °C, the flux ratio of Fe and S is too high, thus there is no enough S to react with Fe, making epitaxial FeS film has a rough morphology with a large number of Fe clusters. Increasing FeS source temperature to 780 °C, Fe clusters have substantially been removed, however hexagonal FeS phase just begins to emerge. Further increasing FeS source temperature to 795 °C, the ratio of hexagonal phase in epitaxial FeS film is significantly increased, meaning that the FeS source temperature is too high for the tetragonal FeS film growth. Through our experiments, we found the optimal temperature of FeS source is 780 °C.

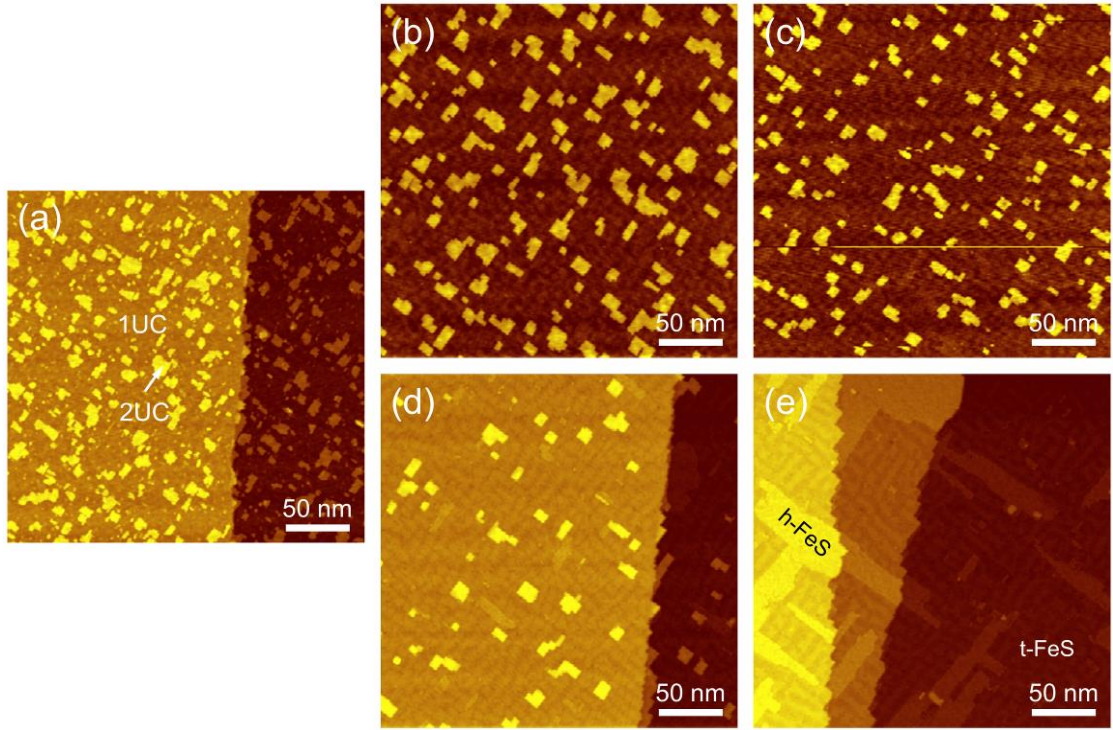

Fig. 3. The effects of post-annealing temperature on the MBE growth of tetragonal FeS film. (a) The STM topographic image of as-grown FeS film grown with optimal substrate and FeS source temperatures (set point: $V_s = 3.0$ V, $I_t = 20$ pA). (b)-(e) The STM topographic images of FeS films annealed in vacuum at temperatures of (b) 350 °C, (c) 400 °C, (d) 450 °C and (e) 500 °C, respectively (set point: $V_s = 3.0$ V, $I_t = 20$ pA).

The MBE growth of high-quality tetragonal FeS film needs to use the combined method of low-temperature growth and high-temperature post-annealing in vacuum, thus the effects of post-annealing temperature were then investigated. Fig. 3a shows the STM topographic images of as-grown FeS film grown with optimal substrate and FeS source temperatures and Fig. 3b-e show the STM topographic images of FeS film annealed in vacuum at temperatures of 350 °C, 400 °C, 450 °C and 500 °C, respectively. With increasing the temperature of post-annealing in vacuum, the edges of FeS islands gradually become more straight, indicating that the crystallinity of epitaxial FeS film becomes better, along with the epitaxial FeS film begins to gradually decompose. Meanwhile, with further increasing the annealing temperature, the hexagonal FeS gradually begins to emerge, and the higher annealing temperature, the larger ratio of hexagonal phase in the epitaxial FeS film. Our experimental results are consistent with that tetragonal FeS is a metastable phase at high temperature. The optimal temperature of post-annealing in vacuum is 400 ~ 450 °C.

Through the above experiments, we have found the optimal growth conditions and achieved the pure tetragonal FeS film with one or two UC thickness on SrTiO$_3$(001) substrate. Fig. 4a shows the typical STM topographic image of epitaxial FeS film grown under the optimal growth condition and Fig. 4b, c show the atomic-resolved STM images of single UC FeS and double UC FeS, respectively. The epitaxial single UC FeS has a tetragonal lattice structure and its in-plane lattice constant is 3.81 Å, which is between the SrTiO$_3$(001) substrate (3.91 Å) and the bulk material of FeS (3.68 Å) [27]. Large expansion of in-plane lattice constant indicates that there is a strong interaction with the SrTiO$_3$(001) substrate, forming a large stress in the epitaxial film. The epitaxial double UC FeS also has a tetragonal lattice structure and its in-plane lattice constant is 3.69 Å, which is almost the same as that of the bulk FeS, indicating the stress has almost been fully released. In addition, the epitaxial double UC FeS has less defects compared with the single UC FeS as shown in Fig. 4b, c.

Then we studied the electronic properties of the epitaxial FeS film through STS. Fig. 4d shows the large-scaled tunneling spectra of epitaxial single UC FeS (blue solid line) and double UC FeS (red solid line). The tunneling spectrum of single UC FeS has an essentially the same shape with that of double UC FeS, but has an upward shift of about 64 mV compared to the double UC FeS. This indicates that there is a significant charge transfer at the interface of the SrTiO$_3$(001) substrate and the epitaxial single UC FeS. Electronic doping effect in the single UC FeS is introduced by the substrate, which was also observed in the single UC FeSe on SrTiO$_3$(001) substrate [7]. Fig. 4e, f show the small-scaled tunneling spectra of epitaxial single UC and double UC FeS. The blue and red solid lines were measured at 4.6 K and 60 mK, respectively. The energy gaps without coherence peaks are both observed near the Fermi level in the single UC and double UC FeS. The energy gap size is almost identical, approximate 1.5 meV at 60 mK. A serious of tunneling spectra of double UC FeS were measured at different spatial locations at 60 mK and are shown in Fig. 4g. The energy gap is spatially homogeneous, but electronic states outside of the gap shows strong spatial variation. The absence of coherence peaks might be related to the incoherent Cooper pairing [33].

To further check this energy gap is a superconducting gap or not, the tunneling spectra of double UC FeS were also measured under the perpendicular magnetic field. As shown in Fig. 4h, the blue solid line was measured without the applied magnetic field and the purple and red solid lines were measured under the applied magnetic fields of 0.05 T and 15 T, respectively. All the spectra were taken at 60 mK. Even under a magnetic field of 15 T, double UC FeS still shows a pronounced energy gap. However, the upper critical field of bulk FeS is only 0.28 T [29]. Therefore, we expect that this observed energy gap should not be a superconducting gap. The absence of interface-enhanced high-temperature superconductivity in this material system is probably due to the quality of epitaxial FeS film is not high enough or intrinsically superconductivity is suppressed under electronic doping or stress for the FeS. The further studies are needed to improve the quality of epitaxial FeS film.

Using the combined method of low-temperature molecular beam epitaxial growth and high-temperature post-annealing in vacuum, we have successfully grown tetragonal FeS film on $SrTiO_3$(001) substrate. Stress and electronic doping of single UC FeS are induced by the $SrTiO_3$(001) substrate. The single UC and double UC FeS have very similar electronic structures and both of them open an energy gap of approximate 1.5 meV at the Fermi level at 4.6 K or below, but no superconductivity is observed in the present experimental condition. It indicates superconducting phase of FeS is much difficult to be reached by the epitaxial growth, more effort and fine controlling is needed in further experiments.

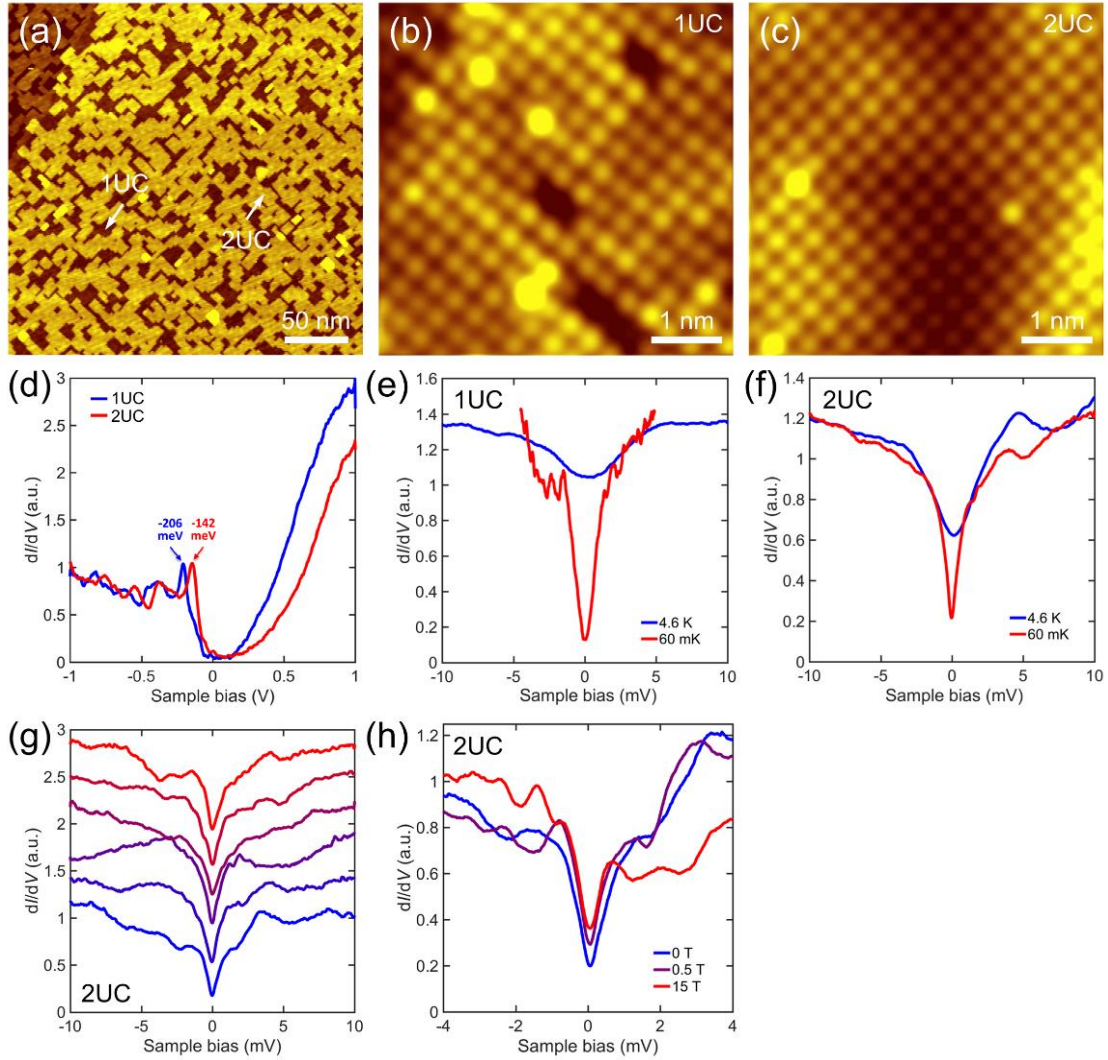

Fig. 4. The lattice structures and electronic properties of epitaxial tetragonal FeS film with one or two UC thickness on SrTiO$_3$(001) substrate. (a) The typical STM topographic image of epitaxial FeS film grown under optimal growth condition (set point: $V_s$ = 3.0 V, $I_t$ = 20 pA). (b), (c) The atomic-resolved STM images of (b) single UC FeS and (c) double UC FeS, respectively (set point: $V_s$ = 0.5 V, $I_t$ = 100 pA). (d) The large-scaled tunneling spectra of epitaxial single UC FeS (blue solid line) and double UC FeS (red solid line) (set point: $V_s$ = 1.0 V, $I_t$ = 100 pA). The spectra were taken at 4.6 K. (e), (f) The small-scaled tunneling spectra of epitaxial (e) single UC FeS and (f) double UC FeS (set point: $V_s$ = 10 mV, $I_t$ = 100 pA). The blue solid lines were measured at 4.6 K and the red solid lines were measured at 60 mK. (g) A serious of tunneling spectra of double UC FeS measured at different spatial locations (set point: $V_s$ = 10 mV, $I_t$ = 100 pA). All the spectra were taken at 60 mK and the curves are offset vertically for clarity. (h) The typical tunneling spectra of double UC FeS under the applied magnetic fields (set point: $V_s$ = 10 mV, $I_t$ = 100 pA). The blue solid line was measured without applied magnetic field. The purple and red solid lines were measured under the applied magnetic fields of 0.05 T and 15 T, respectively. All the spectra were taken at 60 mK.